\documentclass{moriond}

\def\be{\begin{equation}}
\def\ee{\end{equation}}
\def\bea{\begin{eqnarray}}
\def\eea{\end{eqnarray}}

\newcommand{\Photo}{}

\usepackage{subfigure}

\begin{document}
\vspace*{4cm}
\title{Search for exotic physics at BESIII}

\author{Zhi-Jun Li, Zheng-Yun You on behalf of BESIII }

\address{School of Physics, Sun Yat-sen University, Guangzhou 510275, China}

\maketitle\abstracts{
Exotic physics typically encompasses two categories of exotic particles: ``dark" particles beyond the standard model and exotic hadrons within the extended standard model. We present a summary of the recent results on exotic particles at BESIII, which could play a crucial role in the study of exotic physics.
}

\section{Introduction}

The standard model~(SM) has achieved significant success, identifying a series of ``fundamental" particles such as quarks, leptons, photon, Higgs boson, $Z$ boson, $W$ boson, and gluons, which have been extensively studied. However, there are still unresolved puzzles, such as dark matter~(DM), the strong CP problem, the muon $g-2$ anomaly, the fermion mass hierarchy, and more. These puzzles suggest the existence of a dark sector beyond the standard model, potentially containing dark particles and dark mediators that serve as a portal connecting the dark sector with the visible normal sector. The search for these exotic dark particles provides an intriguing avenue for probing new physics~(NP) beyond the SM. Another category of exotic particles, known as exotic hadrons in the extended SM, includes glueballs, multi-quark particles, and hybrids, whose components differ from those of normal hadrons (mesons and baryons). If the mass of these exotic particles falls within the MeV to GeV range, they can be accessed through high-intensity $e^+e^-$ collider experiments.

Beijing Spectrometer III~(BESIII)~\cite{BESIII:2009fln,Huang:2022wuo} is a general-purpose spectrometer for $\tau$-charm physics study in the center-of-mass energy range from 2.0 to 4.7~GeV. BESIII records symmetric $e^+e^-$ collisions provided by the Beijing Electron Positron Collider II~(BEPCII) storage ring~\cite{Yu:2016cof} and has collected large data samples in this energy region~\cite{BESIII:2020nme}, such as 10 billion $J/\psi$ events, 2.7 billion $\psi(2S)$ events, $20~\rm{fb}^{-1}$ data at 3.773 GeV and more than $20~\rm{fb}^{-1}$ data above 4.0 GeV in total. With the world's largest charmonium data samples, significant baryon and meson data samples can also be generated from charmonium decay. Leveraging these extensive data samples at BESIII and mature analytical techniques~\cite{Li:2024pox}, it becomes feasible to investigate exotic particles and new physics beyond the standard model~\cite{Li:2024moj}.

\section{Exotic particles search in BESIII}

\subsection{Search for the invisible massive dark photon}
\begin{figure*}[htbp] \centering
	\setlength{\abovecaptionskip}{-1pt}
	\setlength{\belowcaptionskip}{10pt}

        {\includegraphics[width=0.45\textwidth]{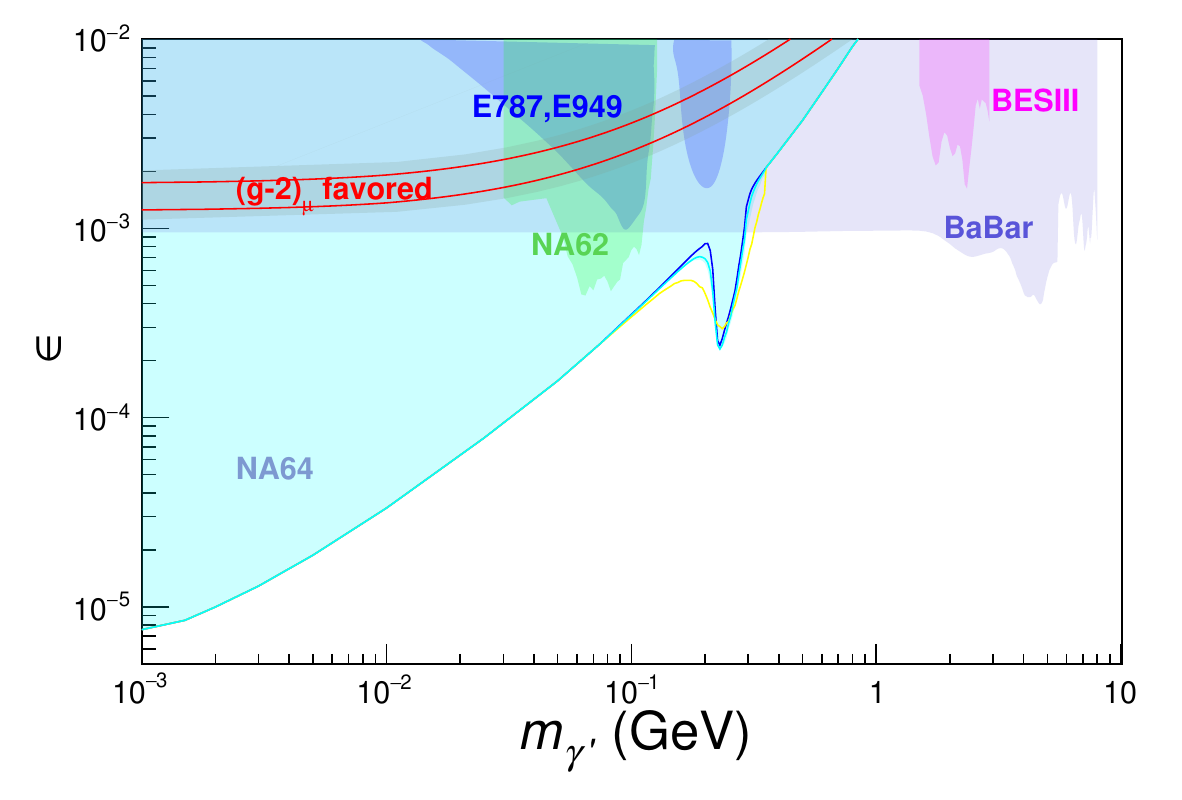}}\\
        
	\caption{
        The 90\% CL UL of the mixing parameter with the dark photon mass less than 10 GeV/$c^2$.
        } 
	\label{fig:mixing}
\end{figure*}

An extra Abelian gauge group may cause the associated gauge boson, the dark photon $\gamma'$~\cite{Holdom:1985ag}. If the symmetry of this extra group is broken spontaneously, the dark photon will be massive. The dark photon always has a kinetic mixing with SM photon, $\frac{1}{2}\epsilon F'_{\mu\nu} F_{\mu\nu}$, where $F_{\mu\nu}$ is the SM photon field strength, $F'_{\mu\nu}$ is the dark photon field strength and $\epsilon$ is a small mixing parameter. Since the SM photon can couple with the SM fermion, this kinetic mixing could be an effective coupling between the SM fermion and the dark photon with coupling term $\mathcal{L}=\frac{e\epsilon}{\sqrt{1-\epsilon^2}}J_{\mu}A'^{\mu}\sim e\epsilon J_{\mu}A'^{\mu}$, where $J_{\mu}$ is the current of ordinary SM matter and $A'_{\mu}$ represents the dark photon~\cite{Fabbrichesi:2020wbt}. The dark photon with mass can be produced in any process by replacing SM photon, such as $e^+e^-\to\gamma\gamma'$ and $J/\psi\to\gamma'\eta^{(')}$, where the dark photon could decay to a lepton pair or be invisible. If $m_{\chi}<m_{\gamma'}/2$ ($\chi$ is the DM particle), the dark photon would predominately decay into a pair of DM particles, which will be invisible at BESIII. Based on $14.9~\rm{fb}^{-1}$ $e^+e^-$ annihilation data at $\sqrt{s}=4.13\sim4.60$ GeV, the invisible massive dark photon is searched for via $e^+e^-\to\gamma\gamma'$, where $\gamma'$ is reconstructed from the visible SM photon $\gamma$ with the relation $E_{\gamma}=\frac{s-m^2_{\gamma'}}{2\sqrt{s}}$. The signals are extracted from the SM photon energy spectrum, and the maximum global significance is $2.2~\sigma$ with no significant signals. The constraint of the mixing parameter can be translated by~\cite{Essig:2009nc}
\begin{eqnarray}
\sigma(e^+e^-\to\gamma\gamma')=\frac{2\pi\alpha^2}{s}\epsilon^2(1-\frac{m^2_{\gamma'}}{s}) \times \left( 
 \left(  1+\frac{2m^2_{\gamma'}/s}{(1-m^2_{\gamma'}/s)^2} \right) \rm{log}\frac{(1+cos\theta_c)^2}{(1-cos\theta_c)^2}  -2cos\theta_c \right),
\label{eq:mixing parameter}
\end{eqnarray}
where $cos\theta_c=0.6$  is the cut for the signal photon polar angle. The  90\% confidence level (CL) upper limit (UL) of the mixing parameter $\epsilon$ is set to be $(1.6\sim5.7)\times10^{-3}$ in the GeV mass region~\cite{BESIII:2022oww}, which is consistent with what already excluded by BaBar, as shown in Fig~\ref{fig:mixing}.
\vspace{-0.0cm}
\vspace{-0.0cm}

\subsection{Search for the massless dark photon}
If the symmetry of the extra Abelian gauge group is unbroken, the dark photon will be massless and be significantly different from the massive dark photon. The massless dark photon has no direct interaction with the SM particle in the dimension-four operator but can be coupled with the SM particle in a higher-dimension operator~\cite{Dobrescu:2004wz}
\begin{eqnarray}
\mathcal{L}_{\rm{NP}}=&\frac{1}{\Lambda^2_{\rm{NP}}} ( C^u_{jk} \bar{q}_j \sigma^{\mu\nu} u_k \tilde{H} + C^d_{jk} \bar{q}_j \sigma^{\mu\nu} d_k H + C^l_{jk} \bar{l}_j \sigma^{\mu\nu} e_k H + h.c. ) F'_{\mu\nu},
\label{eq:dimension-six operator}
\end{eqnarray}
which includes the NP energy scale $\Lambda_{\rm{NP}}$ and dimensionless coefficients $C_{jk}$. In this operator, it has a naturally flavor-violating coupling, such as $cu\gamma'$ coupling. Based on $4.5~\rm{fb}^{-1}$ $e^+e^-$ annihilation data at $\sqrt{s}=4.6\sim4.7$ GeV, the first search for the $cu\gamma'$ coupling is performed through the process $\Lambda_c\to p \gamma'$, where the invisible massless dark photon $\gamma'$ is detected from the recoiling. No significant signals are observed, and the UL of the branching fraction~(BF) is set to be $8.0\times 10^{-5}$ at the 90\% CL~\cite{BESIII:2022vrr}, corresponding to the NP energy scale related parameter $|\mathcal{C}|^2+|\mathcal{C}_5|^2<9.6\times10^{-16} \rm{GeV}^{-2}$, where $\mathcal{C}=\Lambda_{\rm{NP}}^{-2}(C^u_{12}+C^{u*}{12})\nu/\sqrt{8}$, $\mathcal{C}_5=\Lambda_{\rm{NP}}^{-2}(C^u_{12}-C^{u*}{12})\nu/\sqrt{8}$, and $\nu$ is Higgs vacuum expected value~\cite{Su:2020yze}.

\subsection{Search for the muonphilic particle}
\begin{figure*}[htbp] \centering
	\setlength{\abovecaptionskip}{-1pt}
	\setlength{\belowcaptionskip}{10pt}

        \subfigure[]
        {\includegraphics[width=0.32\textwidth]{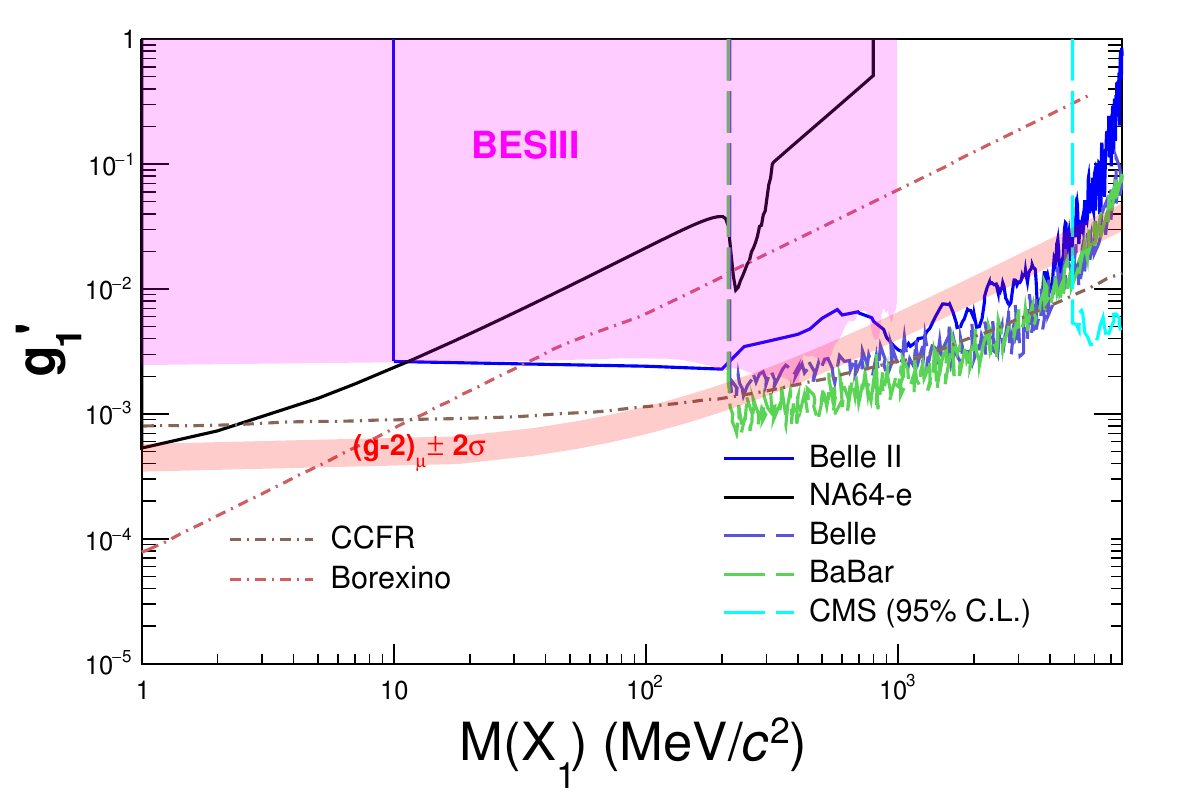}}
        \subfigure[]
        {\includegraphics[width=0.32\textwidth]{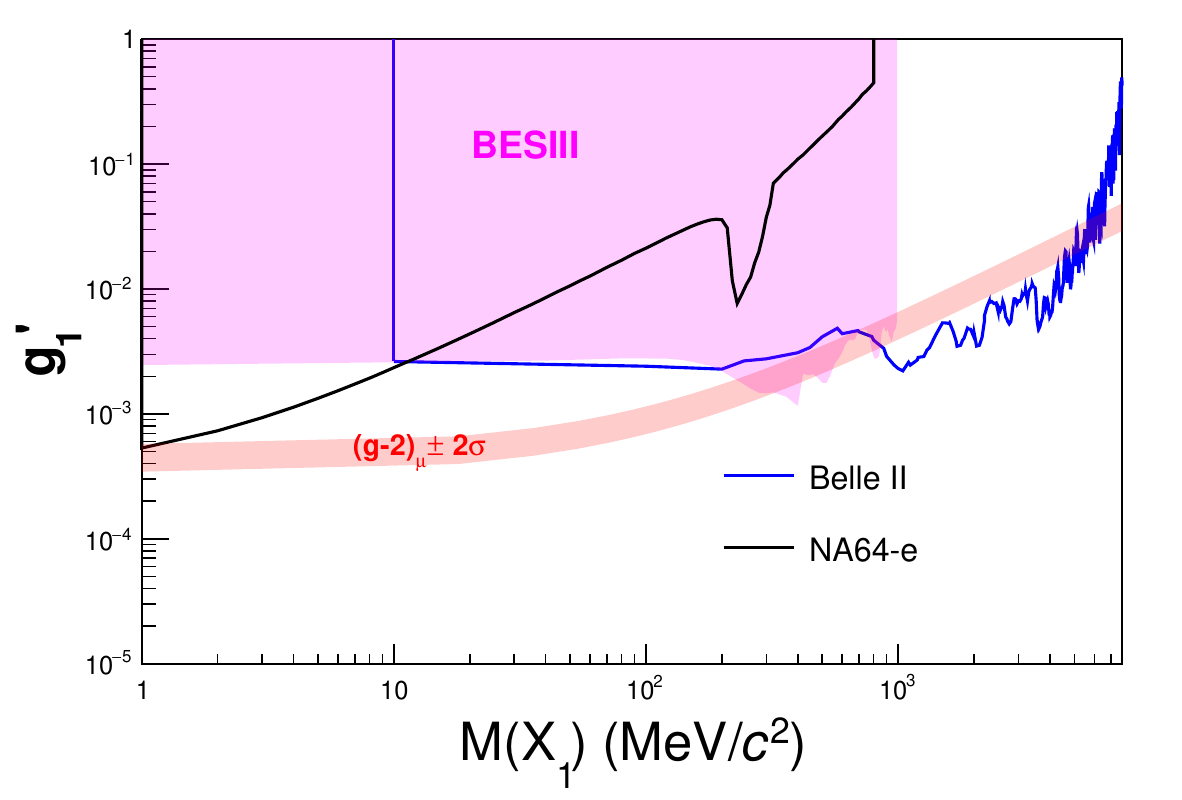}}
        \subfigure[]
        {\includegraphics[width=0.32\textwidth]{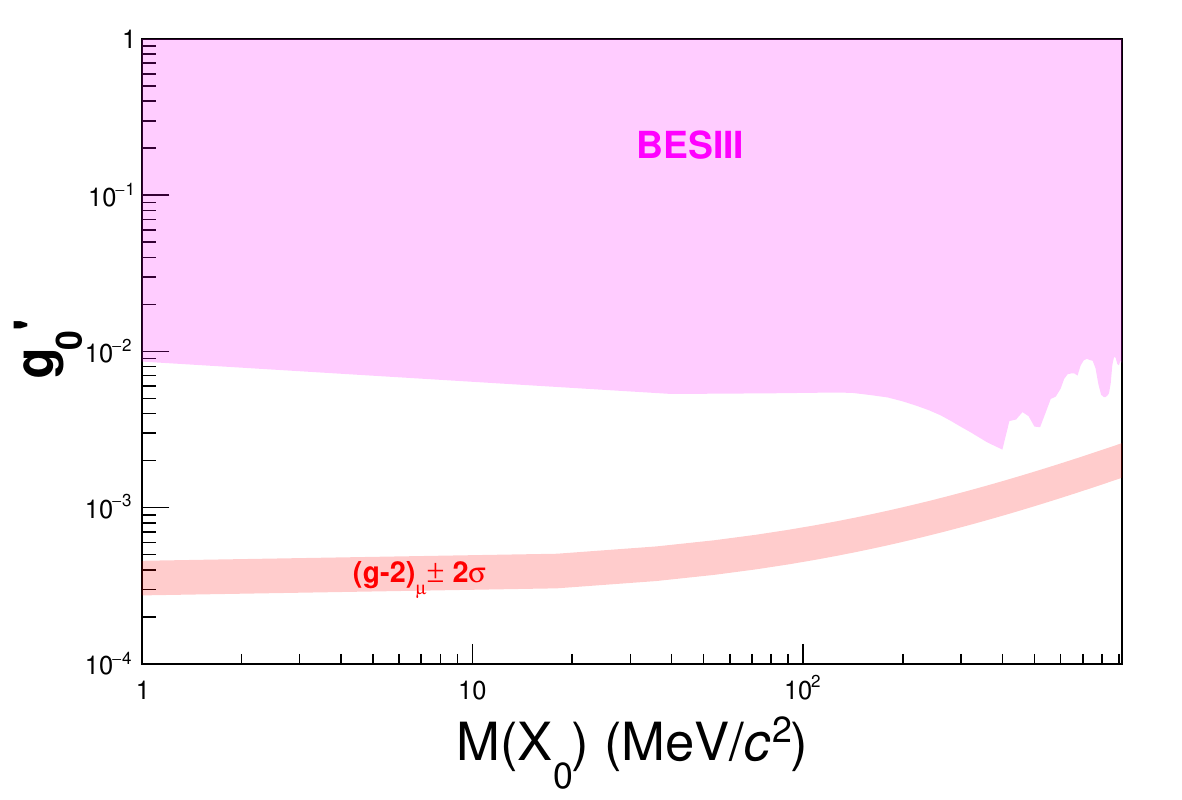}}\\
        
	\caption{
        The constraint of the coupling between muonphilic particle and the SM muon, where (a) is the ``vanilla" model, (b) is the ``invisible" model, and (c) is the ``scalar" model.
        } 
	\label{fig:muonphilic}
\end{figure*}

Similar to the previous dark photon, an extra $U(1)$ group is added as a minimal extension to the SM. There is a $U(1)_{\mathcal{L}_{\mu}-\mathcal{L}_{\tau}}$ model predicts that a new massive scalar boson $X_0$ or vector boson $X_1$ only couples to the second and third generations of leptons ($\mu,~\tau,~\nu_{\mu},~\nu_{\tau}$) with the operator $\mathcal{L}_{\mu}^{\rm{scalar}}=-g'_0 X_0 \bar{\mu}\mu$ and $\mathcal{L}_{\mu}^{\rm{vector}}=-g'_1X_1\bar{\mu}\gamma^{\alpha}\mu$, where $g'_{0,1}$ is the coupling strength. The light muonphilic scalar or vector particles can contribute to the 
muon anomalous magnetic moment and explain the $(g-2)_{\mu}$ anomaly~\cite{Cvetic:2020vkk}. Based on $(8.998\pm0.039)\times10^9$ $J/\psi$ events, the search for the muonphilic particles $X_{0,1}$ is performed via the process $J/\psi\to\mu^+\mu^-X_{0,1}$, where $X_{0,1}$ is invisible in the analysis. Three cases of muonphilic particles are considered. The first one is named the ``vanilla" model with $m_{\chi}>m_{X_1}/2$, and the muonphilic vector particle $X_1$ can only decay to the SM particle pair. $X_1\to\nu\nu$ can be accessible in the invisible final state, where $\mathcal{B}(X_1\to\nu\nu)=33\%\sim100\%$ with different $m_{X_1}$. The second one is named the ``invisible" model with $m_{\chi}<m_{X_1}/2$, and the muonphilic vector particle $X_1$ mainly decay to the DM particle pair with the assumption of $g'_D>>g'_1$. The third one is named the ``scalar" model with the assumption that the muonphilic scalar particle $X_0$ is long-lived or only decay to invisible final states. The invisible muonphilic particle is reconstructed from the recoiling of $\mu^+\mu^-$, and the maximum local significance is 2.5 $\sigma$ at $m(X_{0,1})=720~\rm{MeV}/c^2$. No evidence for signals from $X_{0,1}$ invisible decays are observed, and the coupling constraint at 90\% CL is given in Fig~\ref{fig:muonphilic}. For the ``vanilla" model, Barbar, CMS, and Belle measure with process of $X_1\to\mu^+\mu^-$, while Belle II and BESIII measure with $X_1\to\nu\bar{\nu}$. For the ``invisible" model, BESIII provides better sensitivity in the range of 200 $\sim$ 860 $\rm{MeV}/c^2$. For the ``scalar" model, BESIII gives the first constraint of the “scalar” invisible $X_0$ case.~\cite{BESIII:2023jji}
\vspace{-0.0cm}
\vspace{-0.0cm}

\subsection{Search for the QCD axion}
\begin{figure*}[htbp] \centering
	\setlength{\abovecaptionskip}{-1pt}
	\setlength{\belowcaptionskip}{10pt}

        \subfigure[]
        {\includegraphics[width=0.45\textwidth]{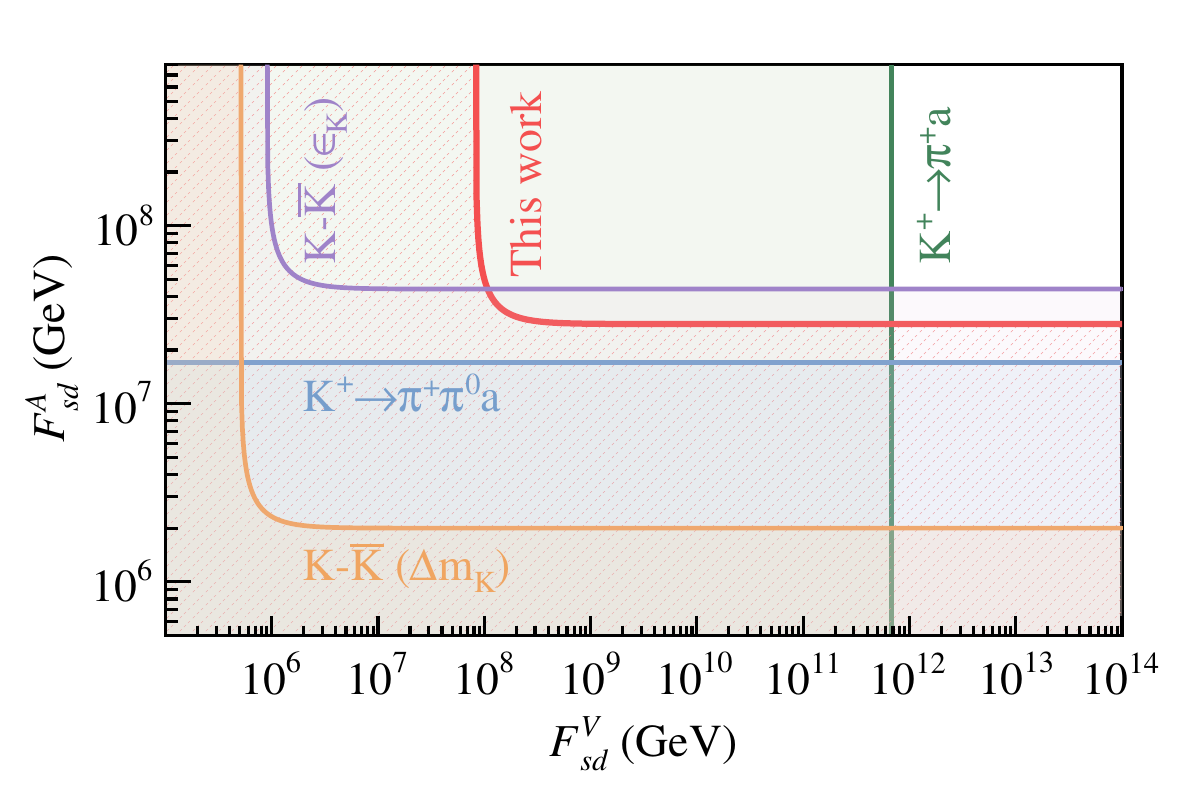}}
        \subfigure[]
        {\includegraphics[width=0.45\textwidth]{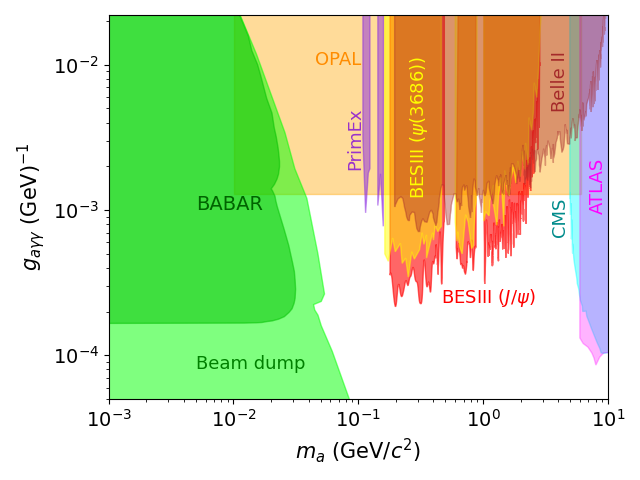}}
        
	\caption{
        (a) The constraint on the effective decay constant of QCD axion associated with $sda$ coupling, where ``this work" is the measurement from $\Sigma^+\to p a$.  (b) The constraint on the ALPs with the mass of ALP less than 10 GeV/$c^2$.
        } 
	\label{fig:axion}
\end{figure*}

The QCD axion~($a$) is originally predicted by the Peccei-Quinn~(PQ) solution to the strong CP problem, which is also an excellent cold dark matter candidate. The mass of the QCD axion is related to the decay constant $f_a$ with the relation of $m_a=5.691(51)~\rm{\mu eV}(\frac{10^{12}~\rm{GeV}}{f_a})$, and the decay constant satisfies $f_a>>10^6$ GeV, making $m_a<\rm{eV}$, which is ``massless" compared to the resolution of BESIII. Such a small mass makes its lifetime larger than the age of the universe and the axion invisible in the detector. The QCD axion could be coupled with the SM fermions with the operator $\mathcal{L}_{a-f}=\partial_{\mu} a \bar{f}_i\gamma^{\mu}(\frac{1}{F^V_{ij}}+\frac{\gamma^5}{F^A_{ij}})$, where $F^V_{ij}$ and $F^A_{ij}$ are the effective decay constants for the vector coupling term and axial coupling term~\cite{MartinCamalich:2020dfe}. If lepton U(1) charges are flavour non-universal, it will have naturally flavour-violating couplings. Based on $(10084\pm44)\times10^6~J/\psi$ events decay to $\Sigma^+\bar{\Sigma}^-$ pair, the QCD axion is searched for through the process of $\Sigma^+\to p a$. A kinematic fit is performed to constrain the invisible axion mass to zero, and the signals are extracted in the energy spectrum of the extra shower in the electromagnetism counter. No significant signals are found, and the UL of $\mathcal{B}(\Sigma^+\to p a)$ is set to be $3.2\times10^{-5}$ at the 90\% CL. The constraint on the effective decay constants can be given by~\cite{MartinCamalich:2020dfe}
\begin{eqnarray}
\Gamma(\Sigma^+\to p a)=\frac{M^3_{\Sigma^+}}{16\pi}\left(  1-\frac{M^2_p}{M^2_{\Sigma^+}} \right) \left(  \frac{(-1)^2}{|F^V_{sd}|^2} +  \frac{(0.34)^2}{|F^A_{sd}|^2} \right).
\label{eq:effective decay constants}
\end{eqnarray}
As shown in Fig~\ref{fig:axion} (a), BESIII provides a competitive constraint on the axial-vectorial effective decay constant $F^A_{sd}>2.8\times10^7$ GeV.~\cite{BESIII:2023utd}
\vspace{-0.0cm}
\vspace{-0.0cm}

\subsection{Search for the axion like particle}
The axion like particles~(ALPs) have the same quantum numbers as the QCD axion but have no strict relation between their couplings and mass, which means that the ALPs can have arbitrary masses and couplings. ALPs can have interaction with fermions, gluon or photons, and the operator of coupling with photons is $\mathcal{L}=-\frac{1}{4}g_{a\gamma\gamma}aF^{\mu\nu}\tilde{F}_{\mu\nu}$ with the coupling strength $g_{a\gamma\gamma}$, leading to the corresponding decay width $\Gamma_{a\to\gamma\gamma}=\frac{g^2_{a\gamma\gamma}m^3_a}{64\pi}$. Taking $g_{a\gamma\gamma}\sim 10^{-4}~\rm{GeV}^{-1}$, $m_a\sim\rm{GeV}/c^2$, the lifetime of ALP will be short in the detector and can be visible via $a\to\gamma\gamma$. Based on $(2.71\pm0.01)\times10^9~\psi(2S)$ events ($\psi(2S)\to\pi^+\pi^-J/\psi$) and another $(10084\pm44)\times10^6~J/\psi$ events ($e^+e^-\to J/\psi$), the ALPs are searched for via the process $J/\psi\to\gamma a,~a\to\gamma\gamma$. The signals are extracted from the $M_{\gamma\gamma}$ spectrum, and the maximum local significance is 2.6 $\sigma$ at $M_a=2208~\rm{MeV}/c^2$ for $\psi(2S)$ data and the maximum global significance is 1.6 $\sigma$ at $M_a=2786~\rm{MeV}/c^2$ for $J/\psi$ data, which means that no evidence for signals from ALPs visible decays are observed. The ALPs-photon coupling constraints are given by~\cite{Merlo:2019anv}
\begin{eqnarray}
\frac{\mathcal{B}(J/\psi\to\gamma a)}{\mathcal{B}(J/\psi\to e^+e^-)}=\frac{m^2_{J/\psi}}{32\pi\alpha} g^2_{a\gamma\gamma} \left(  1-\frac{m^2_a}{m^2_{J/\psi}}  \right)^3.
\label{eq:ALPs}
\end{eqnarray}
As shown in Fig~\ref{fig:axion} (b), BESIII provides an improvement by a factor of 5 over the previous Belle II measurement~\cite{BESIII:2022rzz,BESIII:2024hdv}.

\subsection{Other exotic ``dark" particles search}
\begin{figure*}[htbp] \centering
	\setlength{\abovecaptionskip}{-1pt}
	\setlength{\belowcaptionskip}{10pt}

         {\includegraphics[width=1.0\textwidth]{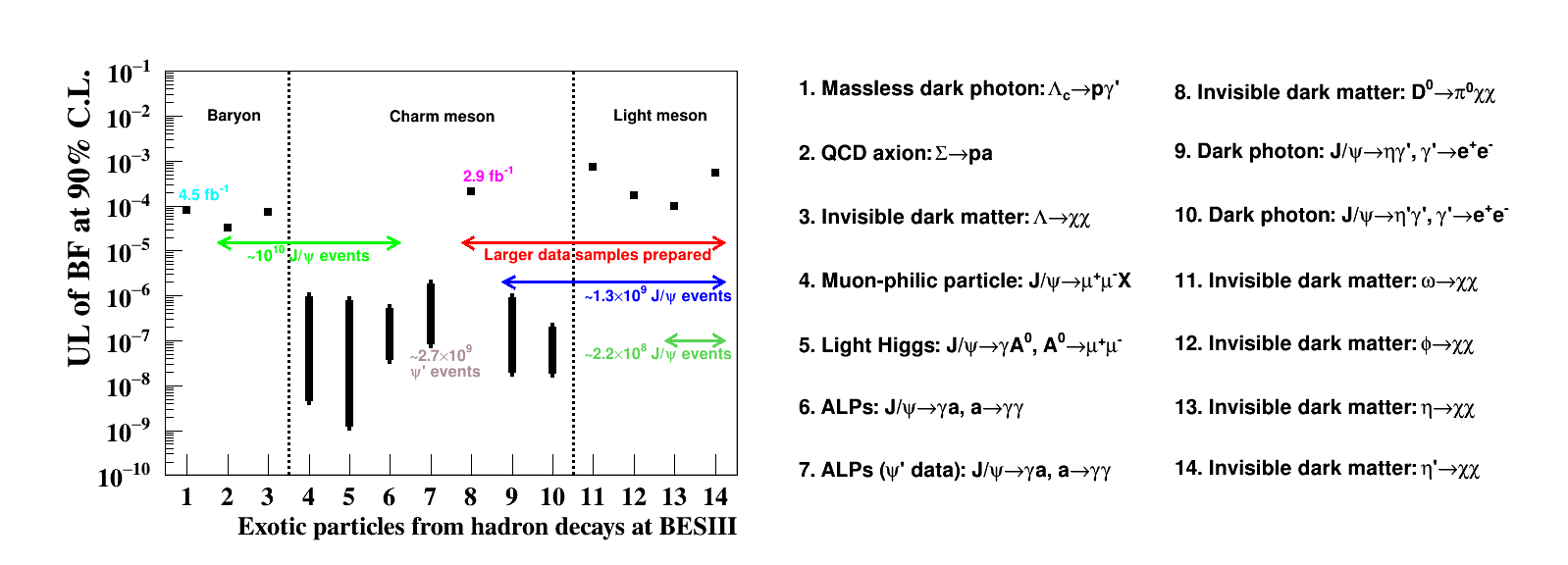}}\\
        
	\caption{
        The summary of the exotic ``dark" particles from hadron decays at BESIII.
        } 
	\label{fig:exotic}
\end{figure*}

There are also some other published exotic ``dark" particle searches at BESIII, such as the full invisible decay $\Lambda/\omega/\phi/\eta/\eta'\to\chi\chi$, semi-invisible decay $D^0\to\pi^0\chi\chi$, visible dark photon search $\gamma'\to e^+e^-$ via $J/\psi\to\gamma'\eta^{(')}$ and light Higgs search $J/\psi\to\gamma A^0,A^0\to\mu^+\mu^-$, which are summarized in Fig~\ref{fig:exotic}.

\vspace{-0.0cm}
\vspace{-0.0cm}

\subsection{Recent glueball studies}
\begin{figure*}[htbp] \centering
	\setlength{\abovecaptionskip}{-1pt}
	\setlength{\belowcaptionskip}{10pt}

        \subfigure[]
        {\includegraphics[width=0.45\textwidth]{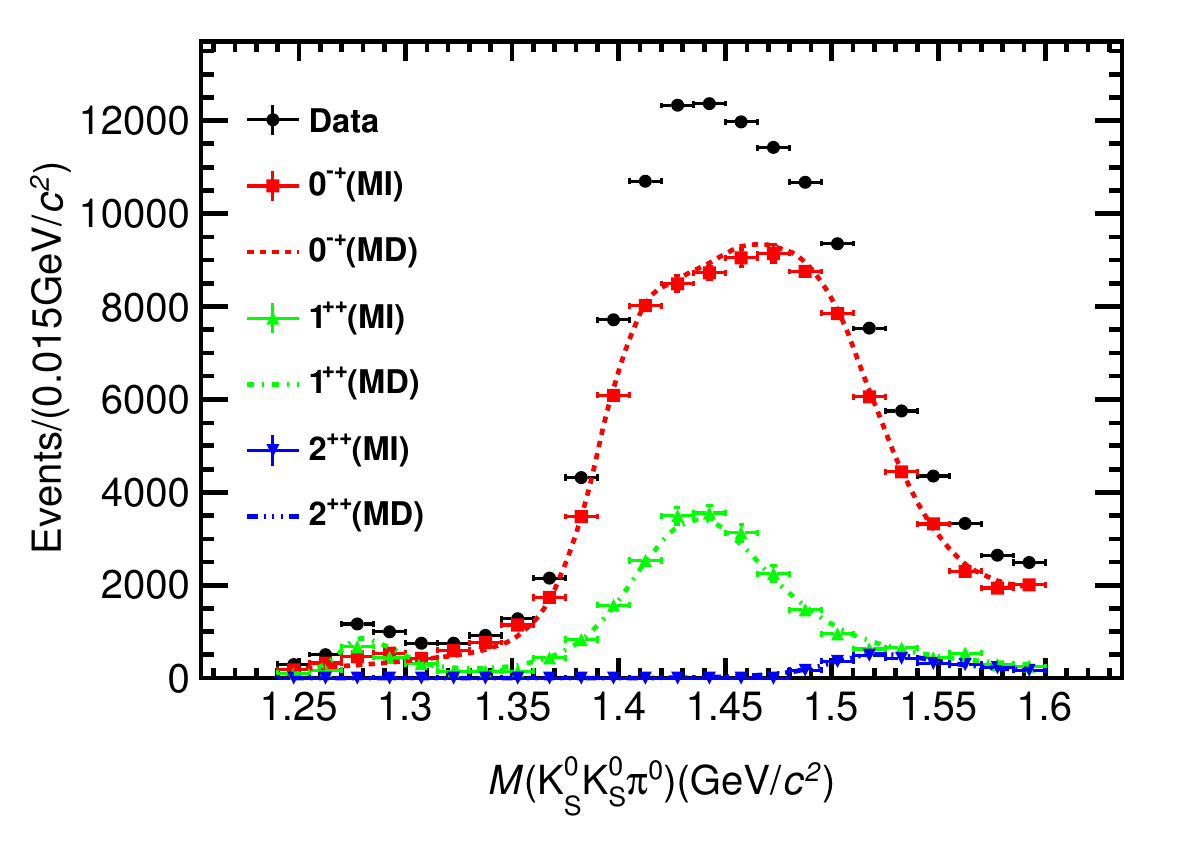}}
        \subfigure[]
        {\includegraphics[width=0.35\textwidth]{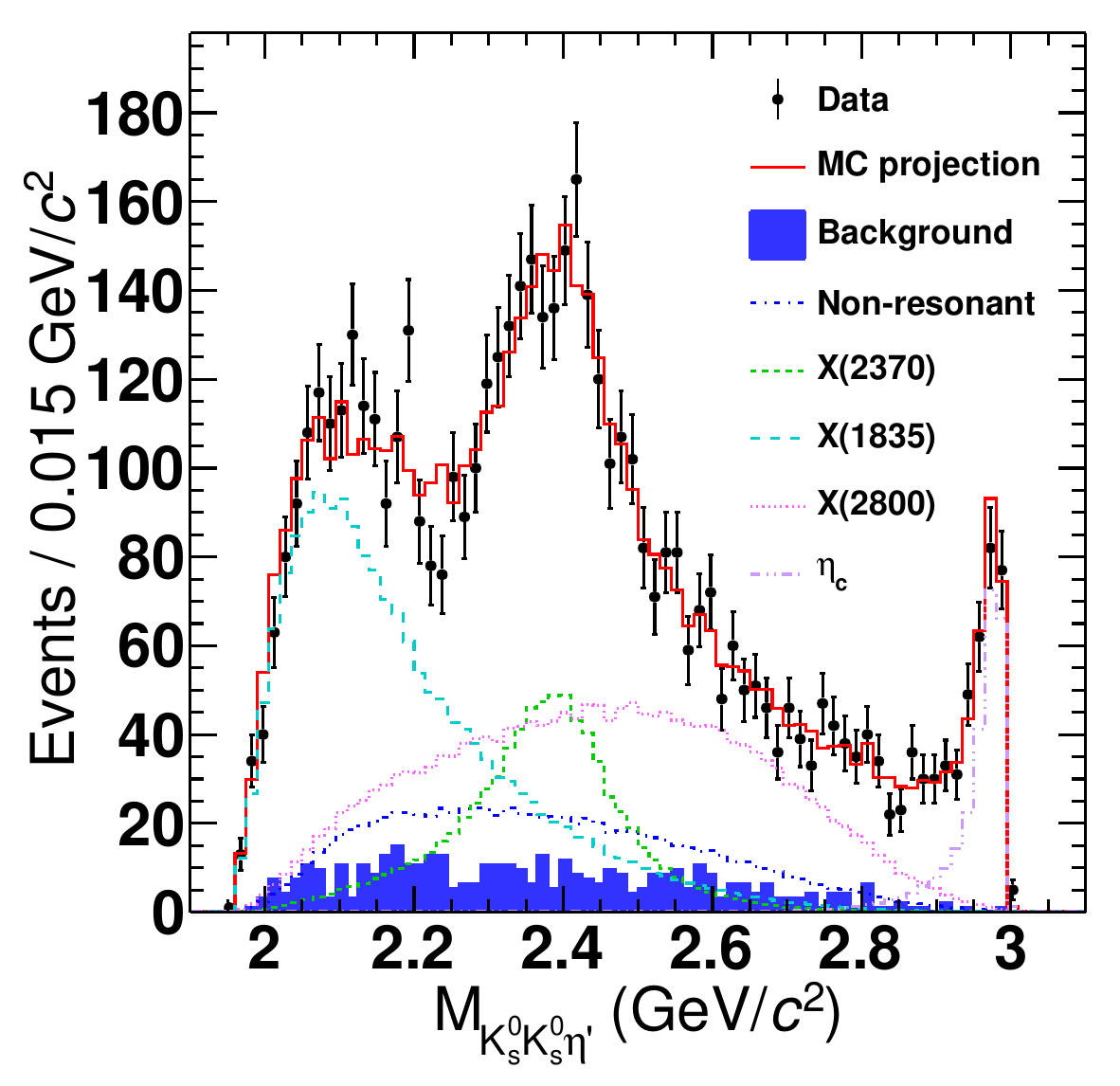}}
        
	\caption{
        (a) The PWA of $J/\psi\to \gamma K_S K_S \pi^0$, where MI is Mass independent PWA, MD is Mass dependent PWA and the results from these two methods are consistent. (b) The PWA of $J/\psi\to\gamma K_S K_S \eta'$.
        } 
	\label{fig:glueball}
\end{figure*}

$J/\psi$ radiative decays provide gluon rich environment and an ideal place for glueball study. A partial wave analysis~(PWA) of $J/\psi\to \gamma K_S K_S \pi^0$ is performed at BESIII, and two pseudo-scalar states are observed, $\eta(1405)$ and $\eta(1475)$~\cite{BESIII:2022chl}, as shown in Fig~\ref{fig:glueball} (a). In the Quark model prediction, there is only one pseudo-scalar meson near 1.4 GeV, and $\eta(1475)$ is regarded as the first radial excitation of $\eta'$ while $\eta(1405)$ may be a pseudo-scalar glueball candidate. This measurement could be an important input for $0^{-+}$ glueball. In the LQCD prediction, the mass of $0^{-+}$ glueball is in the range of 2.3 $\sim$ 2.4 $\rm{GeV}/c^2$. Some theorists attempt to explain $\eta(1405)$/$\eta(1475)$ using a single state~\cite{Wu:2011yx} , while others propose more than one pole~\cite{Nakamura:2022rdd}.

\vspace{-0.0cm}
\vspace{-0.0cm}

Another pseudo-scalar glueball candidate is $X(2370)$. $X(2370)$ was firstly observed in $J/\psi\to\gamma \pi^+\pi^-\eta'$~\cite{BESIII:2010gmv}, and a new PWA of $J/\psi\to\gamma K_S K_S \eta'$ is performed at BESIII, as shown in Fig~\ref{fig:glueball} (b), resulting that $M_{X(2370)}=(2395\pm11~^{+26}_{-0.84})~\rm{MeV}/c^2$, $\Gamma_{X(2370)}=(188~^{+18}_{-17}~^{+124}_{-33})~\rm{MeV}/c^2$, and $J^{PC}_{X(2370)}=0^{-+}$, where the spin-parity of $X(2370)$ is determined to be $0^{-+}$ for the first time~\cite{BESIII:2023wfi}. The measured mass and spin-parity of the $X(2370)$ are all consistent with the predictions of a pseudo-scalar glueball.

\section*{Acknowledgments}

This work is supported in part by National Key Research and Development Program of China under Contracts Nos. 2023YFA1606000, 2020YFA0406400, 2020YFA0406300; Joint Large-Scale Scientific Facility Funds of the National Natural Science Foundation of China (NSFC) and Chinese Academy of Sciences (CAS) under Contracts Nos. U1932101.

\end{document}